\documentclass [12pt]{article}
\usepackage {amssymb}
\usepackage {amsmath}
\usepackage[english]{babel} 
\usepackage{mathdots}
\usepackage[dvips]{graphicx} 

\begin{document}

\begin{center}
\Large\textbf{A comprehensive model of high-concentration phosphorus diffusion in silicon}
\\[2ex]
\normalsize
\end{center}

\begin{center}
\textbf{O.~I. Velichko}
\end{center}

\begin{center}
\bigskip

Full Professor of Physics department, Belarusian state university of informatics and radioelectronics, 6, P. Brovka Str., Minsk, 220013 Republic of Belarus \textbf{}

velichkomail@gmail.com\textit{ }

\end{center}

\noindent \textbf{}

\noindent \textbf{Abstract: }A comprehensive model of high-concentration phosphorus diffusion has been developed and simulation of phosphorus diffusion from a constant source (phosphosilicate glass) at a temperature of 890$^{\circ}$C for 14.25 min\textbf{ }has been carried out\textbf{.} Such doping processes are widely used in manufacturing modern solar cells. The proposed model combines the ideas of the drift of silicon self-interstitials in the field of elastic stresses with the concept of the formation of negatively charge clusters of impurity atoms. The calculated phosphorus concentration profile is in good agreement with the experimental one.

\bigskip


\bigskip

\section{Introduction}

In contrast to manufacturing integrated microcircuits and semiconductor devices, doping with phosphorus is widely used to form the emitters of solar cells. This is due to the capability of high-concentration phosphorus layers for gettering undesirable iron-type impurities \cite{Phang-2011,Park-2012}. In this connection, it makes sense to develop a comprehensive model which would describe the high-concentration phosphorus diffusion and allow adequate simulation of doping processes with the aim of designing solar cells.

\begin{figure}[ht]
\centering
\includegraphics[width=4.50in,height=3.2in]{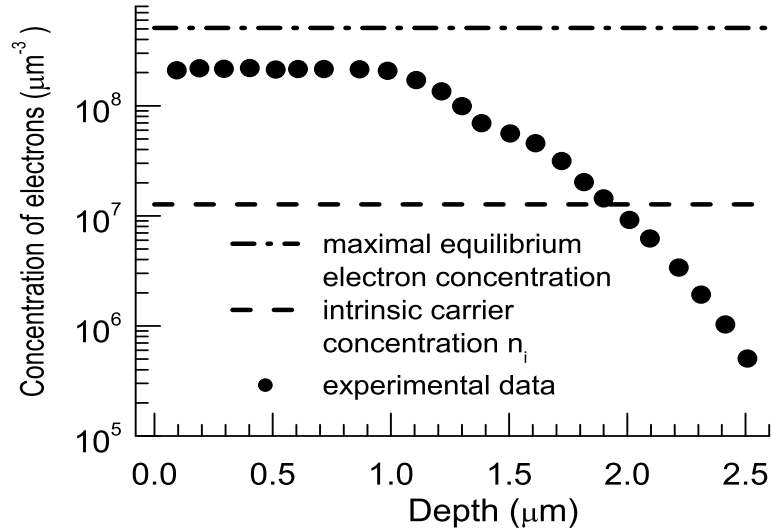}
\caption{Electron density vs. depth for high-concentration phosphorus diffusion under condition of a constant impurity concentration on the surface of a semiconductor. The dash-dotted line represents the maximal equilibrium concentration of charge carriers \cite{Solmi-2001}. The dashed line is the concentration of intrinsic charge carriers $n_{i} $. The experimental data are taken from \cite{Panteleev-1977} for diffusion at a temperature of 1050$^{\circ}$C for 30 min. \label{fig:Panteleev1050}}
\end{figure}

Among the impurities used for the doping of silicon, phosphorus diffusion is characterized by the most intricate behavior. For example, the distribution of the concentration of charge carriers in a silicon substrate formed by high-concentration thermal diffusion of phosphorus is characterized by a ``plateau'' with a constant concentration of electrons. On the other hand, the concentration profile of substitutionally dissolved phosphorus atoms is characterized by a ``kink'', where the impurity distribution convexity changes its sign to the opposite, and by an extended ``tail'' in the region of low impurity concentration. In Fig.~\ref{fig:Panteleev1050}, the experimental profile of electron concentration measured in \cite{Panteleev-1977} is presented. This figure illustrates the formation of the above-mentioned ``plateau'', ``kink'', and the ``tail'' in the course of high-concentration thermal diffusion of phosphorus atoms from a constant source on the surface of a semiconductor. Another specific feature of the high-concentration diffusion of phosphorus is the enhanced redistribution of boron, gallium, phosphorus, and arsenic atoms in the previously formed buried layers \cite{Lecrosnier-1979,Pichler-2004}.

A detailed analysis of the basic models of the high-concentration diffusion of phosphorus is presented in \cite{Pichler-2004,Velichko-1987,Velichko-1999,Velichko-2013}. It is assumed in the latest models that the diffusion of phosphorus atoms in the ``tail'' region occurs through the formation of the ``phosphorus atom--silicon self-interstitial'' pairs \cite{Velichko-1987,Velichko-2013,Mulvaney-1987,Orlowski-1988,Uematsu-1997}. Unlike the diffusion mechanisms considered earlier, the mechanism of the formation of mobile ``phosphorus atom--silicon self-interstitial'' pairs makes it possible to explain many of the laws governing the processes of high-concentration phosphorus diffusion. Thus, the assumption that the self-interstitial distribution is nonuniform makes it possible to explain the specific form of the distribution of phosphorus atoms in the course of diffusion from a constant source \cite{Velichko-1987,Mulvaney-1987,Orlowski-1988}. This nonuniform distribution can be formed both by the absorption of silicon self-interstitials in the highly doped region \cite{Velichko-1987} and as a consequence of dissociation of pairs in the low concentration region \cite{Mulvaney-1987}. The assumption of the generation of nonequilibrium silicon interstitials makes it possible to identically explain the processes of enhanced diffusion of phosphorus atoms in the ``tail'' region and the enhanced impurity redistribution in the buried layers \cite{Velichko-1987}. Also, the dissociation of the complexes and diffusion of generated silicon self-interstitials allow explaining the phenomenon of supersaturation of the bulk of a semiconductor with self-interstitials. However, these models fail to explain the disappearance of the distinctive features of high-concentration phosphorus diffusion with decrease in the internal elastic stresses in the case of phosphorus and germanium codiffusion.

\bigskip

It is interesting to note that a considerable spatial nonuniformity of point defect distribution should be observed in the near-surface region to provide the formation of the  ``kink'', and ``tail'' on the impurity concentration profile within the limits of the models   \cite{Velichko-1987,Mulvaney-1987,Orlowski-1988,Uematsu-1997}. It means that the average migration length of silicon self-interstitials should be small enough. On the other hand, it is usually supposed that the average migration length of silicon self-interstitials is significantly larger (approximately 30 $\mu$m in accordance with the experimental data describing the reduction in the enhancement of impurity diffusion on increase of the distance from the surface \cite{Lecrosnier-1979}). It means that a considerable decrease in the average migration length of silicon self-interstitials is observed in the region between the surface and the ``kink''. Not rejecting the possibility of such phenomenon, it is worthwhile to consider another reason for the formation of a strong nonuniform distribution of point defects.

For example, it was supposed in \cite{Velichko-1999,Velichko-2013,Fedotov-2004} that internal elastic stresses are generated in the vicinity of the surface due to the lattice deformation. If the drift velocity of silicon self-interstitials in the field of elastic stresses is directed to the bulk of a semiconductor, the near-surface region is depleted of point defects, and a strong nonuniform distribution of self-interstitials can be formed without regard for the large migration length of these defects. Indeed, a dramatic compression of the crystal lattice will be observed in the heavily doped region in high-concentration phosphorus diffusion due to the difference in the covalent radii of the phosphorus (0.110 nm) and silicon (0.117 nm) atoms \cite{Ormont-1982}. The drift of the nonequilibrium silicon self-interstitials in the field of elastic stresses leads to the supersaturation of the bulk of a semiconductor with self-interstitials and, consequently, to the enhanced diffusion of impurity atoms in the buried layers. The phosphorus concentration profiles calculated in \cite{Velichko-2013,Fedotov-2004} on the basis of these assumptions agree well with experimental data, including the formation of the ``kink'' and ``tail'' regions. The phenomena of oversaturation of the bulk of a semiconductor with self-interstitials \cite{Schaake-1984} and enhanced impurity redistribution in the buried layers were also explained. Unfortunately, in \cite{Fedotov-2004}, the cluster formation was not taken into account, whereas in \cite{Velichko-2013}, the model of the formation of neutral phosphorus clusters was used. Therefore, only the initial stage of high-concentration phosphorus diffusion was simulated where cluster formation is negligible.

\section{Model}

For comprehensive modeling high-concentration phosphorus diffusion with a quantitative description of the ``plateau'', characterized by a constant electron density, it is necessary to combine the model of phosphorus diffusion \cite{Velichko-2013,Fedotov-2004} with the model of the formation of negatively charged phosphorus clusters \cite{Velichko-2008}. It means that a set of equations describing high-concentration phosphorus diffusion, namely: (i) equation of diffusion of phosphorus atoms \cite{Velichko-2013,Fedotov-2004}; (ii) approximation of local charge neutrality \cite{Velichko-2008} for the concentration of electrons normalized to the concentration of intrinsic charge carriers $n_{i}$; (iii) equation of vacancy diffusion, \cite{Velichko-2013,Fedotov-2004}  and (iv) equation of the diffusion of silicon self-interstitials \cite{Velichko-2013,Fedotov-2004} must be combined with the equation describing cluster formation \cite{Velichko-2008}

\begin{equation}\label{Cluster concentration}
C^{T} =C+C^{AC} ,
\end{equation}

\begin{equation}\label{Clustering}
C^{AC} =K\tilde{C}_{Cl} \, \chi ^{(m-z_{Cl} )} C^{m} .
\end{equation}

Here $C$ and\textit{ }$C^{AC} $ are respectively the concentrations of phosphorus atoms in the substitutional position and incorporated into clusters; $C^{T} $ is the total phosphorus concentration; $m$ is the number of substitutionally dissolved phosphorus atoms, participating in cluster formation, and $z_{Cl} $ is the charge of this cluster expressed in units of the elementary charge; the parameter $K$ has a constant value depending on the temperature of diffusion and the function $\tilde{C}_{Cl} $ describes the influence of nonuniform distributions of defects participating in the reaction of cluster formation on the concentration of clustered impurity atoms \cite{Velichko-2008}.

It follows from the values of the enhanced impurity diffusion in ``buried'' layers \cite{Lecrosnier-1979} that the concentration of nonequilibrium silicon self-interstitials beyond the phosphorus-doped layer is many times higher than the thermally equilibrium concentration of these point defects. This means that in a highly doped layer, silicon self-interstitials are generated. For example, such generation may occur when phosphorus atoms are built into the crystal lattice of silicon, as has been assumed in \cite{Faney-1984}. Furthermore, at a high phosphorus concentration, precipitates of silicon phosphoride SiP \cite{Ravi-1984} or clusters of phosphorus atoms are formed \cite{Velichko-2008}. The quasichemical reactions providing cluster formation or dissolution may be accompanied by the generation of silicon self-interstitials. In turn, the generated silicon self-interstitials find themselves in the field of elastic stresses, which provide their drift beyond the region of high phosphorus concentration. To find the distribution of the effective drift velocity of the silicon self-interstitials, the lattice deformation and distribution of the internal elastic stresses in the doped layer must be calculated. However, the distribution of these stresses depends not only on the distribution of the phosphorus atoms but also on the distribution of other components of the defect-impurity system of the crystal, namely, on the vacancies, silicon self-interstitials, clusters or precipitates of phosphorus atoms. The solution of this problem is extremely difficult and is not the aim of the present work. Therefore, we approximate the distribution of the effective drift velocity of silicon self-interstitials by the following semiempirical dependence \cite{Velichko-2013}:

\begin{equation}\label{Drift velocity}
{\rm v}_{{\rm x}}^{{\rm I}} (x)=\left\{\begin{array}{c} {\begin{array}{l} {{\rm v}_{\max }^{\, I} \; ,\quad x<x^{*} } \\ {} \end{array}} \\ {{\rm v}_{\max }^{\, I} \exp \left[-\frac{\left(x-x^{*} \right)}{2\Delta R_{pst}^{2} } \right]\, ,\quad x\ge x^{*} } \end{array}\right.  .
\end{equation}

Here ${\rm v}_{\max }^{\, I} \; $ is the maximum value of the drift velocity of silicon self-interstitials; $x^{*} $ is the characteristic dimension of the stress region; $\Delta R_{pst}^{2} $ is the dispersion of the distribution of the drift velocity of point defects. The parameter $x^{*} $ characterizes the depth of the influence of the internal elastic stresses and may be related to the characteristic value of the impurity concentration $C^{*} =C(x^{*} )$, which determines the position of the inflection point on the phosphorus profile. It is worth noting that the semiempirical dependence proposed above reflects all characteristic properties of distribution of elastic stresses in the layer doped with phosphorus and have the shape similar to the distribution of electron density.

\noindent

\section{Results of calculations}

\begin{figure}[ht]
\centering
\includegraphics[width=4.50in,height=3.2in]{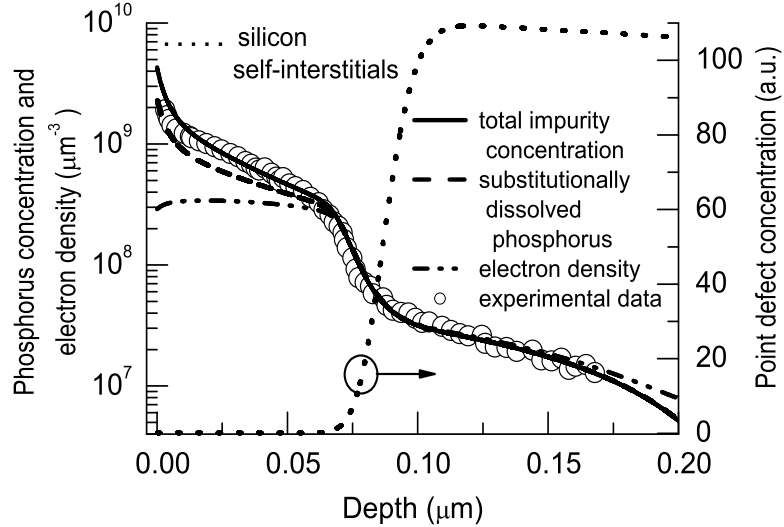}
\caption{Calculated phosphorus concentration profile (solid curve) in the case of high-concentration diffusion from a constant source at 890$^{\circ}$C for 14.25 min. Dash curve is the concentration of substitutionally dissolved phosphorus atoms, dash-dot-dot curve describes distribution of electron density, and dotted curve represents the distribution of silicon self-interstitials in a neutral charge state at the end of thermal treatment, normalized to the thermally equilibrium concentration of these particles. Circles are the experimental data according to \cite{Siddique-2018}. \label{fig:Phosphorus890}}
\end{figure}

Typical results of modeling the high-concentration phosphorus diffusion on the basis of the model proposed are presented in Fig.~\ref{fig:Phosphorus890}. The data of \cite{Siddique-2018} were used to compare the results of calculation and experiment. In \cite{Siddique-2018}, phosphorus was introduced into the substrate using a phosphor-oxy-trichloride (POCl${}_{3}$) source. High-concentration phosphorus doping was carried out using the following gases: 99.99999\% pure POCl${}_{3}$, 99.9\% pure O${}_{2}$, and 99.998\% pure N${}_{2}$. A phosphosilicate glass (PSG) is formed on the surface of a silicon wafer during the treatment in this gas mixture. According to the data of \cite{Siddique-2018}, this layer of PSG can be considered as a diffusion source with a constant impurity concentration on the semiconductor surface. The profile of the total concentration of phosphorus atoms measured by SIMS for the diffusion at a temperature of 890$^{\circ}$C for 14.25 min is also presented in Fig.~\ref{fig:Phosphorus890}.

It is clearly seen from this figure that the numerical calculation of high-concentration  phosphorus diffusion is in good agreement with experimental data. The following values of the parameters describing the transport of impurity atoms and intrinsic point defects were used in modeling: the intrinsic concentration of charge carriers $n_{i} $ = 4.61$\mathrm{\times}$10${}^{6}$ $\mu$m${}^{-}$${}^{3}$, the maximal equilibrium concentration of charge carriers in phosphorus layers for this temperature of diffusion $n_{e}$ = 3.419$\mathrm{\times}$10$^{8}$ $\mu$m${}^{-3}$ \cite{Solmi-2001}, the intrinsic diffusivity of phosphorus $D_{i}$ = 6.5698$\mathrm{\times}$10${}^{-}$${}^{8}$ $\mu$m${}^{2}$/s \cite{Pichler-2004}. The value of the relative contribution of the indirect interstitial mechanism in the transport of impurity atoms under low-concentration diffusion $f_{i} $ = 0.27 was taken from \cite{Collard-1986}. It follows from this value of $f_{i} $ that $D_{i}^{E} $\textit{ }= 4.8$\mathrm{\times}$10${}^{-}$${}^{8}$ $\mu$m${}^{2}$/s and $D_{i}^{F} $\textit{ }= 1.771$\mathrm{\times}$10${}^{-}$${}^{8}$ $\mu$m${}^{2}$/s. Here $D_{i}^{E} $ and $D_{i}^{F}$ are respectively the intrinsic diffusivities of impurity atoms due to the mechanism of the formation, migration, and dissociation of vacancy--impurity complexes and due to the mechanism of the formation, migration, and dissociation of ``phosphorus atom--silicon self-interstitial'' pairs. The value of the average migration length of silicon self-interstitials $l_{i}^{I} $\textit{ }= 30 $\mu$m was selected in accordance with the experimental data that describe the reduction in the enhancement of diffusion with increasing distance from the surface \cite{Lecrosnier-1979}. It is supposed that nonequilibrium silicon self-interstitials are generated at the surface of a semiconductor and (DP${}_{2}$)${}^{2-}$ clusters are formed in the high-concentration region ($m$= 2 and $z_{Cl} $ = -2) \cite{Velichko-2008}.

Other values of the parameters of the diffusion process were determined from the condition of the best agreement between the calculated distribution of the total phosphorus concentration and experimental data. For example, the concentration of substitutionally dissolved phosphorus atoms at the semiconductor surface $C_{S} $\textit{ }= 2.3$\mathrm{\times}$10${}^{9}$ $\mu$m${}^{-3}$, the parameter describing a cluster formation $K\tilde{C}_{Cl} $ = 2.45$\mathrm{\times}$10${}^{-17}$ $\mu$m${}^{3}$, the values of the parameters determining the concentration dependence of diffusivity due to vacancy--impurity complexes\textbf{ }${\rm \beta }_{1}^{E} $ = 0.42, ${\rm \beta }_{2}^{E} $ = 1.0$\mathrm{\times}$10${}^{-5}$ and the values of the parameters determining the concentration dependence of diffusivity due to ``phosphorus atom--silicon self-interstitial'' pairs ${\rm \beta }_{1}^{F} $ = 0.91, ${\rm \beta }_{2}^{F} $ = 0. The character concentration $C^{*} $ related to the formation of the inflection point and of the ``kink'' on the phosphorus concentration profile is equal to 3.25$\mathrm{\times}$10${}^{8}$ $\mu$m${}^{-3}$. We used the following values describing the defect subsystem of silicon substrate: the normalized concentration of silicon self-interstitials in a neutral charge state at the surface of a semiconductor $\tilde{C}_{S}^{I\times } $ = 0.11 a.u., the maximum value of the effective drift velocity of silicon self-interstitials is ${{\rm v}_{\max }^{\, I}  \mathord{\left/{\vphantom{{\rm v}_{\max }^{\, I}  d_{i}^{I} }}\right.\kern-\nulldelimiterspace} d_{i}^{I}} $= 370.0 $\mu$m${}^{-}$${}^{1}$, and $\Delta R_{pst} $= 0.016 $\mu$m. Here $d_{i}^{I}$ is the intrinsic diffusivity of silicon self-interstitials. The value of the normalized concentration of neutral vacancies at the surface of a semiconductor $\tilde{C}_{S}^{V\times}$ was chosen to be 0.45 a.u., and the average migration length of this species $l_{i}^{V}$ = 0.007 $\mu$m. Using these values provides a formation of nonuniform distribution of neutral vacancies in the vicinity of the surface that results in the formation of the peak of phosphorus concentration adjacent to the surface.

\section{Analysis of simulation results}

It is clear from Fig.~\ref{fig:Phosphorus890} that the numerical simulation of high-concentration phosphorus diffusion, carried out on the basis of the model developed in this work, is in good agreement with experimental data and describes all regularities of the impurity concentration profile. In contrast to the model developed in \cite{Velichko-2013,Fedotov-2004}, the model presented here explains the saturation of electron density under conditions of high-concentration phosphorus diffusion. Moreover, taking into account the formation of negatively charge clusters \cite{Velichko-2008}, allows us to remove all limits of the model \cite{Velichko-2013,Fedotov-2004} and simulate different processes of high-concentration phosphorus diffusion, including high-temperature long-time treatments and diffusion with very high surface concentration. As is seen from Fig.~\ref{fig:Phosphorus890}, the concentration of silicon self-interstitials in the bulk of a semiconductor exceeds the thermally equilibrium concentration approximately 100 times, i.e., the intensity of impurity diffusion due to the ``phosphorus atom--silicon self-interstitial'' pairs grows in the ``tail'' region an identical number of times. It is worth noting that the value of the contribution of the indirect interstitial mechanism used in calculation was $f_{i} $ = 0.27. It means that a nearly 28-fold enhancement of the process of diffusion of phosphorus atoms occurs in the bulk of a semiconductor compared to the thermally equilibrium low-concentration diffusion.

\section{Conclusions}

The model of high-concentration phosphorus diffusion has been developed, which combines the ideas of the drift of silicon self-interstitials in the field of elastic stresses with the concept of the formation of negatively charged clusters of impurity atoms. Taking into account the formation of nonuniform distributions of vacancies and silicon self-interstitials in a neutral charge state, this comprehensive model explains all specific features of high-concentration phosphorus diffusion. The results of modeling of phosphorus diffusion from a constant source at a temperature of 890$^{\circ}$C for 14.25 min are in good agreement with experimental data. The calculations show that the proposed model explains the entire shape of the phosphorus concentration profile, including the formation of the inflection point and the ``kink'', where the convexity of the impurity distribution changes its sign to the opposite, as well as the extended ``tail'' in the region of low impurity concentration. The proposed model also explains the oversaturation of the bulk of a semiconductor with silicon self-interstitials and saturation of electron density in the high-concentration region. The contradiction between the large migration length of self-interstitials (approximately 30 $\mu$m) and the formation of strong nonuniform distribution of these point defects is obviated. Absorption of vacancies by the ``phosphosilicate glass---silicon'' interface allows explaining the formation of a peak of phosphorus concentration adjacent to the surface of a semiconductor. Thus, the numerical calculations curried out in this investigation confirm the adequacy of the proposed model of high-concentration phosphorus diffusion.

\end{document}